# Evaluating Metal-Organic Precursors for Focused Ion Beam Induced Deposition through Solid-Layer Decomposition Analysis

B.R. Jany[a*], K. Madajska[b], A. Butrymowicz-Kubiak[b], F. Krok[a], I.B. Szymańska[b]

[a]Marian Smoluchowski Institute of Physics, Faculty of Physics, Astronomy and Applied Computer Science, Jagiellonian University, Lojasiewicza 11, 30348 Krakow, Poland
[b]Faculty of Chemistry, Nicolaus Copernicus University in Toruń, Gagarina 7, 87-100 Toruń, Poland

[*]Corresponding author e-mail: benedykt.jany@uj.edu.pl

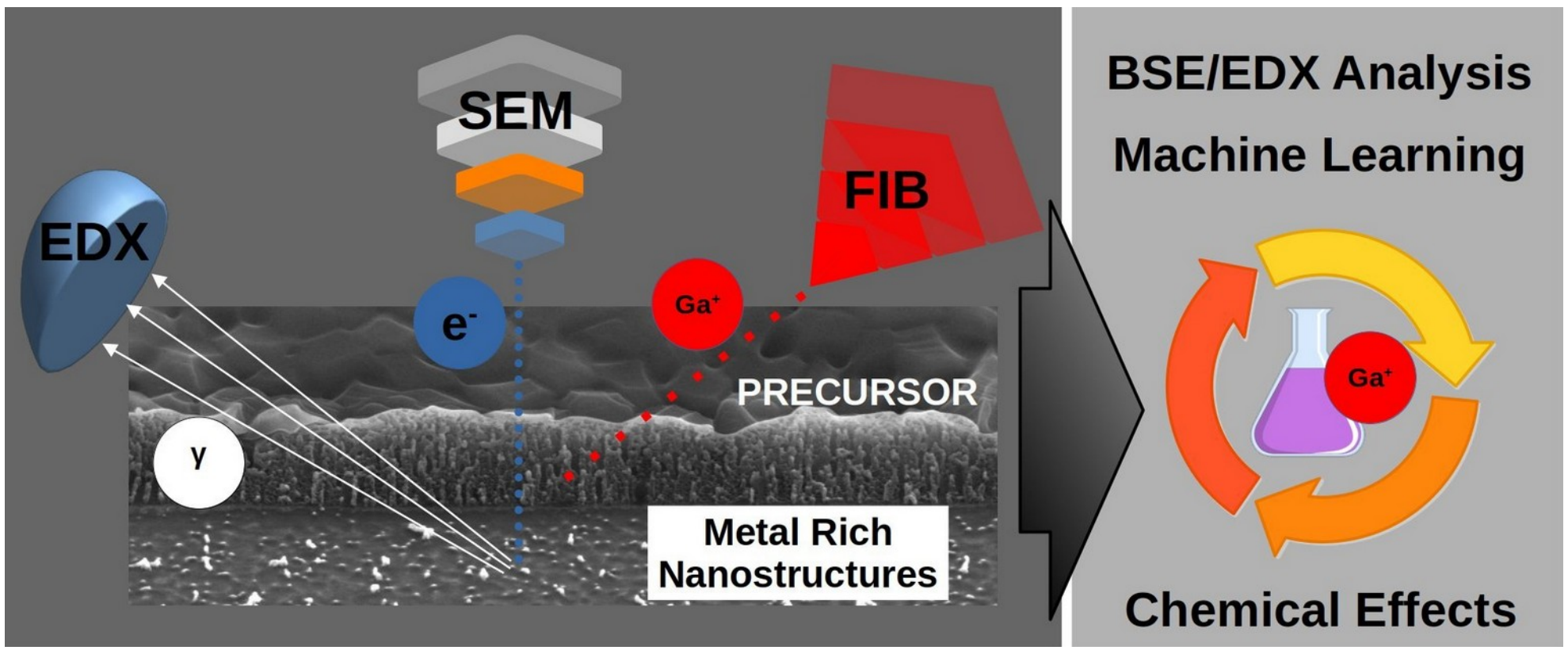

**Abstract**

The development of modern metal deposition techniques like Focused Ion/Electron Beam Induced Deposition FIBID/FEBID relies heavily on the availability of metal-organic precursors of particular properties. To create a new precursor, extensive testing under specialized gas injection systems is required along with time-consuming and costly chemical analysis typically conducted using scanning electron microscopes. This process can be quite challenging due to its complexity and expense. Here, the response of new metal-organic precursors, in the form of supported thick layers, to the ion beam irradiation is studied through analysis of the chemical composition and morphology of the resulting structures. This is done using SEM BSE/EDX along with Machine Learning data processing techniques. This approach enables a comprehensive fast examination of precursor decomposition processes during FIB irradiation, and provides valuable insights into how the precursor's composition influences the final properties of the metal-rich deposits. Although solid-layer irradiation differs from gas-phase deposition, we think that our method, can be employed to optimize pre-screen and score new potential precursors for FIB applications by significantly reducing the time required and conserving valuable resources.

## Introduction

The various nanomanufacturing technologies, such as optical/electron-beam lithography, nanoimprint lithography (NIL), atomic layer deposition (ALD), chemical mechanical polishing (CMP), or laser nanopatterning, allow for the fabrication of nanostructures and devices. Their significant disadvantage is the problem with the production of high-resolution 3D deposits[1,2,3,4]. In comparison, focused electron/ion beam induced deposition (FEBID/FIBID) techniques enable two- and three-dimensional nanostructures of precisely defined shape and size of 5−10 nm[5,6,7,8,9,10,11]. The high lateral resolution can be achieved by controlling the electron/ion beam in terms of the location and duration of the pulse. In the case of FEBID and FIBID, the volatile precursor molecules are dosed by a gas injection system (GIS) onto a substrate surface, where they are adsorbed and are decomposed by a focused electron or ion beam with keV energy. Commercial FIBID instruments typically use $Ga^+$ ions, but it is also possible to use others such as $He^+$, $Ne^+$, $Ar^+$, or $Xe^+$[12,13,14,15]. The resulting FEBID/FIBID deposits are widely used for repairing photolithographic masks and printing or modifying integrated circuits. In addition, they are applied for the fabrication or modification of cantilevers in AFM and scanning optical near-field (SNOM) microscopy, and as plasmonic materials[16,17,18,19,20]. FEBID/FIBID techniques combine the advantages of direct-write lithographic processes, for example, high spatial resolution, site-specific, maskless, and resistance with the flexibility of depositing materials on non-planar surfaces[4,5,15].

The FIBID method has several advantages compared to the FEBID technique in depositing thin films on substrates. Firstly, ions generate more secondary electrons on the substrate surface than electrons, leading to faster deposition growth (around 100 times). Secondly, FIBID deposits have higher metal content and lower resistivity compared to FEBID. However, there are some disadvantages to FIBID, such as the larger size of noble gas and metal ions that penetrate to smaller depths in solids and result in significant beam-induced substrate defects (e.g., Ga atoms implementation). Additionally, material growth is required to compete with the FIB milling process[4,10]. The use of ions instead of electrons, like in FEBID, offers several benefits, including enhanced film quality and adhesion, better control over the growth process, and greater flexibility in material selection (the ability to deposit a variety of different materials). The usage of ions opens new possibilities for materials development and applications[21],[22].

Until now the development of FEBID has relied on precursors used for chemical vapour deposition (CVD), a thermally driven process. However, these kinds of precursors were not optimized for the electron- and ion-driven FEBID and FIBID processes[4,5,15]. The important class of FEBID tested compounds for group 11 elements have been β-diketonates and carboxylates. These compounds were used previously in the chemical vapour deposition method and β-diketonates are the most common CVD precursors (high purity films up to 99 at.%[2,5,23]. However, in FEBID, silver(I) carboxylates, in contrast to β-diketonates, result in high metal content in the deposits. Recent research using $[Ag_2(\mu\text{-}O_2CR)_2]$, where R = $CF_3$, $C_2F_5$, $C_3F_7$, $^tBu$, $C(Me)_2Et$) showed that these carboxylates can be dissociated via focused electron beam yielding deposits with satisfying metal content (purity up to 76 at.% Ag). However, for the copper(II) carboxylate $[Cu_2(\mu\text{-}O_2CC_2F_5)_4]$, the fabricated materials have only up to 23 at.%[5,13,24,25,26,27]. This shows that the electron beam induced decomposition is influenced by the ligand and also by the coordination centre.

Here one has to mention that testing new metal-organic precursors for the use in FEBID/FIBID is a tedious time consuming task which requires costly experimental (non commercial) GIS system. The primary objective of this precursor testing is to optimize deposition parameters, specifically targeting high metal content (favoring minimal gallium from FIB source) and reduced ion currents. Several tests of many different new precursors have to be done before deciding which compound is the most promising one. Therefore, we used in our studies of copper(II) and silver(I) carboxylate complexes such as non-fluorinated pivalate $[Cu_2(\mu\text{-}O_2C^tBu)_4]_n$[28,29], perfluorinated pentafluoropropionates $[Cu_2(\mu\text{-}O_2CC_2F_5)_4]$[30], $[Ag_2(\mu\text{-}O_2C_2F_5)_2]$[31] and the heteroligand complex with the same carboxylate and pentafluoropropamidine $[Cu_2(NH_2(NH=)CC_2F_5)_2(\mu\text{-}O_2CC_2F_5)_4]$[32], as new potential precursors for the applications in Focus Ion Beam Induced Deposition (FIBID) using gallium ions.

Here, we present a pathway for unraveling how chemical composition of metal-organic precursors affects their decomposition when irradiated with FIB in the form of supported layers. A decomposition process of the studied layer was quantitatively monitored by SEM BSE electrons analysis. For each studied precursor an optimal ion fluence was determined, defined as the ion fluence at which the sputtering of the formed metal-rich structures becomes the dominant process, exceeding the rate of precursor decomposition and material buildup. While sputtering occurs throughout ion irradiation, this point marks the transition beyond which further irradiation leads primarily to material removal rather than the structure growth. The structures formed at “the optimal ion fluence” were examined by SEM EDX together with Machine Learning based hyperspectral data processing, which uses non-negative matrix decomposition (NMF) method to extract the EDX signal of structures from the ones of substrate. As already showed this type of analysis greatly enhances the applicability of SEM EDX for the analysis of nanostructures[33]. Finally, we determined the quantitative chemical composition of the formed metal-rich deposits (structures). While the decomposition of precursor layers presents a significant challenge compared to gas-phase deposition, driven by differing mechanisms, our approach of layer analysis offers crucial insights into the fundamental physics of metal deposition from metal-organic precursors. Furthermore, we believe that our methodology could be effectively utilized as a valuable tool for precursor screening. The use of a new compound as a FIBID/FEBID precursor necessitates a series of preliminary tests to confirm volatility and sensitivity to secondary electrons. Refining the conditions for efficient precursor delivery via the GIS system and its subsequent decomposition under ion beam influence requires testing in a difficult-to-access experimental reactor. Given the time- and cost-intensive nature of analyzing the morphology and composition of the deposits, we propose a method/approach to minimize studies within the experimental reactor and identify promising potential precursors.

**Testing pathway and methods used**

The proposed approach for effectively testing new metal-organic precursors involves a series of steps that are crucial to ensure accurate and comprehensive results. These stages include:

1. Deposition of the precursor onto a Si(111) substrate through sublimation process using previously established parameters [23,26,34]. This step allows for precise control of the thickness of the precursor layer on the substrate.
2. Performing SEM imaging of the growth layers, which provide detailed information about the surface structure and composition of the precursor layer. They are essential for understanding how the precursor are decomposed under the following FIB irradiation and what kind of morphology is developed for the finally formed metal rich structures.
3. Analyzing BSE (Backscattered Electron) images of the evolving surface morphology at successive stages of FIB irradiation in order to determine “the sputtering point” - the threshold ion fluence at which the sputtering becomes the dominant process over structure growth, leading to the erosion of the formed metal-rich deposits. This threshold provides insight into the precursor’s resistance to ion bombardment and is crucial for assessing its stability and reactivity under processing conditions.
4. Collecting SEM EDX (Energy Dispersive X-ray) hyperspectral data, which involves acquiring multiple x-ray spectra from different points of the final sample morphology. This step allows for a more detailed analysis of the chemical composition and distribution of elements within the irradiated sample.
5. Decomposing SEM EDX hyperspectral data using advanced algorithms to separate and identify individual components within the irradiated sample. This process is essential for obtaining accurate and reliable information about the chemical composition of developed deposits.

6. Determining the chemical composition of the developed structures using EDX ZAF technique, which is a high-resolution analytical method that can provide elemental information at the nano level. This step ensures precise identification and quantification of all elements present in the grown structures.
7. The final step involves examining the chemical composition of the resulted precursor layers and coupling it with ion beam parameters to score the precursor usability. This stage is crucial for determining the potential applications and limitations of new metal-organic precursors in various fields.

In the following, each of these steps will be explored in greater detail, providing a more in-depth understanding of the proposed pathway for successful testing of the new potential FIBID precursors.

**The fabrication of the precursor thin layer**

The metal-organic precursors films for the FIB/SEM experiments were deposited by sublimation using a glassware sublimation apparatus. The Si(111) wafer was placed in the special holder on the cold finger of the apparatus. The process was performed under a pressure of $10^{-2}$ mbar, at the following temperatures: 418 K − $[Cu_2(\mu\text{-}O_2C^tBu)_4]_n$[23] (**1**), 393 K − $[Cu_2(NH_2(NH{=})CC_2F_5)_2(\mu\text{–}O_2CC_2F_5)_4]$[26] (**2**), 413 K − $[Cu_2(\mu\text{-}O_2CC_2F_5)_4]$ (**3**), and $[Ag_2(\mu\text{–}O_2CC_2F_5)_2]$ (**4**) (Fig. 1). The conditions for depositing the layers of the compounds (**1**) and (**2**) were previously determined[23,26]. In a similar way, the layers of the complexes (**3**) and (**4**) were prepared. The grown layer compositions were checked by IR spectroscopy before the electron beam irradiation. IR spectra were registered with a Vertex 70V spectrometer (Bruker Optik, Leipzig, Germany) using a single reflection diamond ATR unit (400–4000 $cm^{-1}$). IR spectra of the obtained layers and native compounds $[Cu_2(\mu\text{-}O_2C^tBu)_4]_n$ (**1**), $[Cu_2(NH_2(NH{=})CC_2F_5)_2(\mu\text{-}O_2CC_2F_5)_4]$ (**2**), $[Cu_2(\mu\text{-}O_2CC_2F_5)_4]$ (**3**), and $[Ag_2(\mu\text{-}O_2CC_2F_5)_2]$ (**4**) are presented in Fig. 1b). The spectra showed characteristic $\nu_{as}$COO and $\nu_s$COO bands of bonded carboxylate ligands (**1−4**), as well as exhibited $\nu_{as}NH_2$, ν(=NH), $\delta NH_2$, and νN=C−N bands of coordinated amidine ligand for the complex (**2**), see Fig. 1b, Table S1, confirming the formation of the suitable layers of the studied complexes.

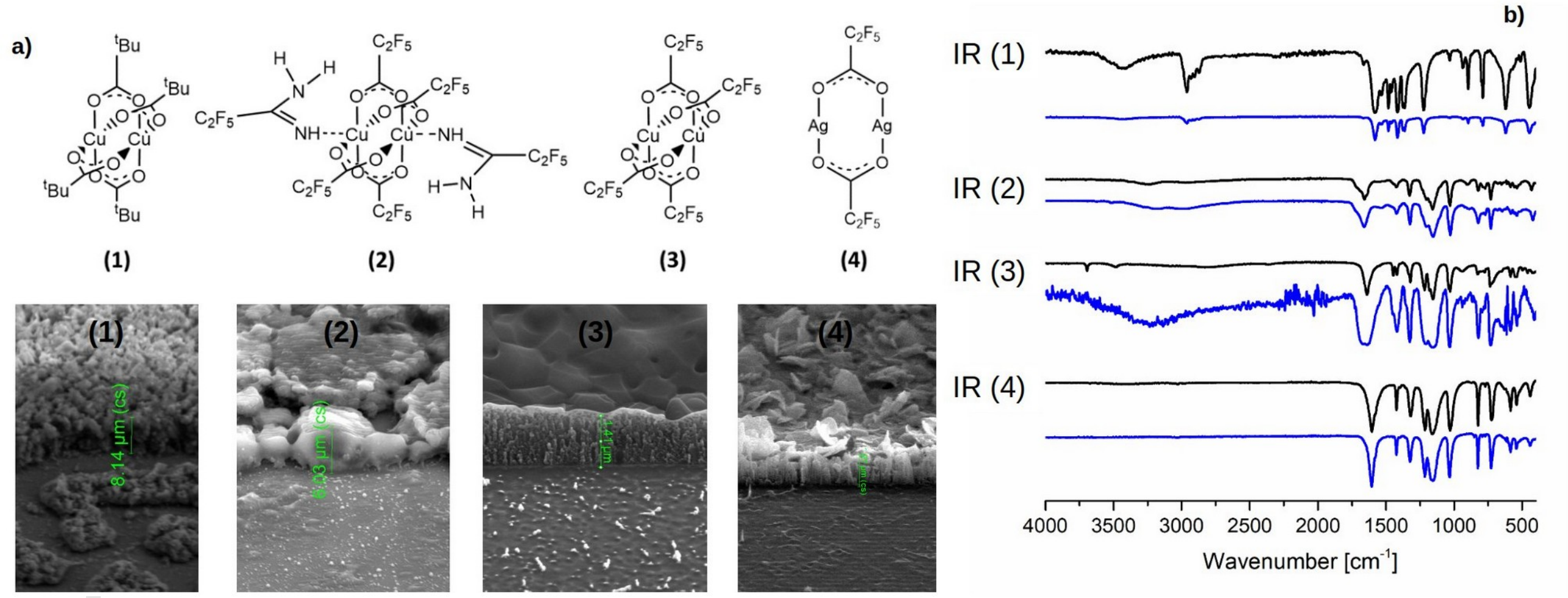

*Figure 1: SEM SE images showing thickness and morphology of used precursors layers (for tested complexes (1−4) as shown in a). b) Infrared spectra before (black) and after sublimation on a silicon wafer (blue) for the compounds: (1) – $[Cu_2(\mu\text{-}O_2C^tBu)_4]_n$, (2) − $[Cu_2(NH_2(NH{=})CC_2F_5)_2(\mu\text{-}O_2CC_2F_5)_4]$, (3) − $[Cu_2(\mu\text{-}O_2CC_2F_5)_4]$, and (4) − $[Ag_2(\mu\text{-}O_2C_2F_5)_2]$ (p = $10^{-2}$ mbar).*

The Scanning Electron Microscopy (SEM) operating in Secondary Electrons (SE) mode provided clear and precise images of the deposited layers, as depicted in Fig.1a). Also, with the help of SEM measurements, the thickness of the grown layers was determined, yielding 8.14 µm for the film no. (1), 6.03 µm for no.(2), 1.41 µm for no.(3), and 1.07 µm for film no.(4). The experiments were conducted using a Dual Beam SEM/FIB Microscope Quanta 3D FEG, manufactured by the FEI. The microscope is equipped with a gallium FIB (Focused Ion Beam) and an EDAX Ametek SDD EDX detector setup.

**FIB/SEM irradiation experiments and EDX chemical composition quantification**

In these FIB irradiation/SEM imaging experiments, a 30 keV energy beam was employed for raster scanning over 50 µm by 50 µm square area with a dwell time of 200 nanoseconds. The ion beam current, as well as the duration of the experiments, were precisely adjusted (within a range of 1 to 10 nA for the ion current, and 10 s to 10 min the irradiation) in order to achieve the optimal ion fluence necessary for the decomposition of the entire precursor layer. Time-dependent changes in the morphology of the irradiated films were tracked using SEM BSE signal, which is directly proportional to the average atomic number Z. The BSE morphology changes during Ga FIB experiments for the precursor (**4**) are presented in Fig. 2a. The initial layer consists of grain-like structures, the blurred BSE contrast indicates that the layer has a rather homogeneously distributed chemical composition, and only a few low-contrast grooves are visible. During the ion irradiation, the BSE contrast already strengthens strongly in the very early stages of bombardment. An increasing number of ("grid") dark contrast grooves separating elongated island-like features become more apparent. The precursor layer underwent decomposition leading to the development of surface features enriched with metallic element of the primarily film.

The quantitative changes are presented in Fig. 2b which shows mean BSE signal intensity, acquired while imaging the surface structures formed, as a function of the ion-beam irradiation time. It is seen that at the initial stages of irradiation, the BSE signal rapidly rises (in comparison to the reference i.e. not irradiated sample). Next one sees increase of the metal content, and the precursor decomposes. Finally, the BSE signal rapidly drops; all the precursors already decomposed into metal rich phase, the sputtering of the metal phase has started to dominate, and the layer is getting thinner and thinner and finally is sputtered out (as in Fig.1 b)-e) ). From this dependence, one can determine the optimal sputtering point, i.e. the maximal ion dose to decompose the given precursor layer without dominant sputtering, after we stop the experiments. In Fig. 2c BSE image morphologies are shown for four studied precursors (**1**−**4**). The BSE morphology for the initial (reference) material is presented together with the BSE morphology for the precursors after Ga FIB experiments for the optimal sputtering point.

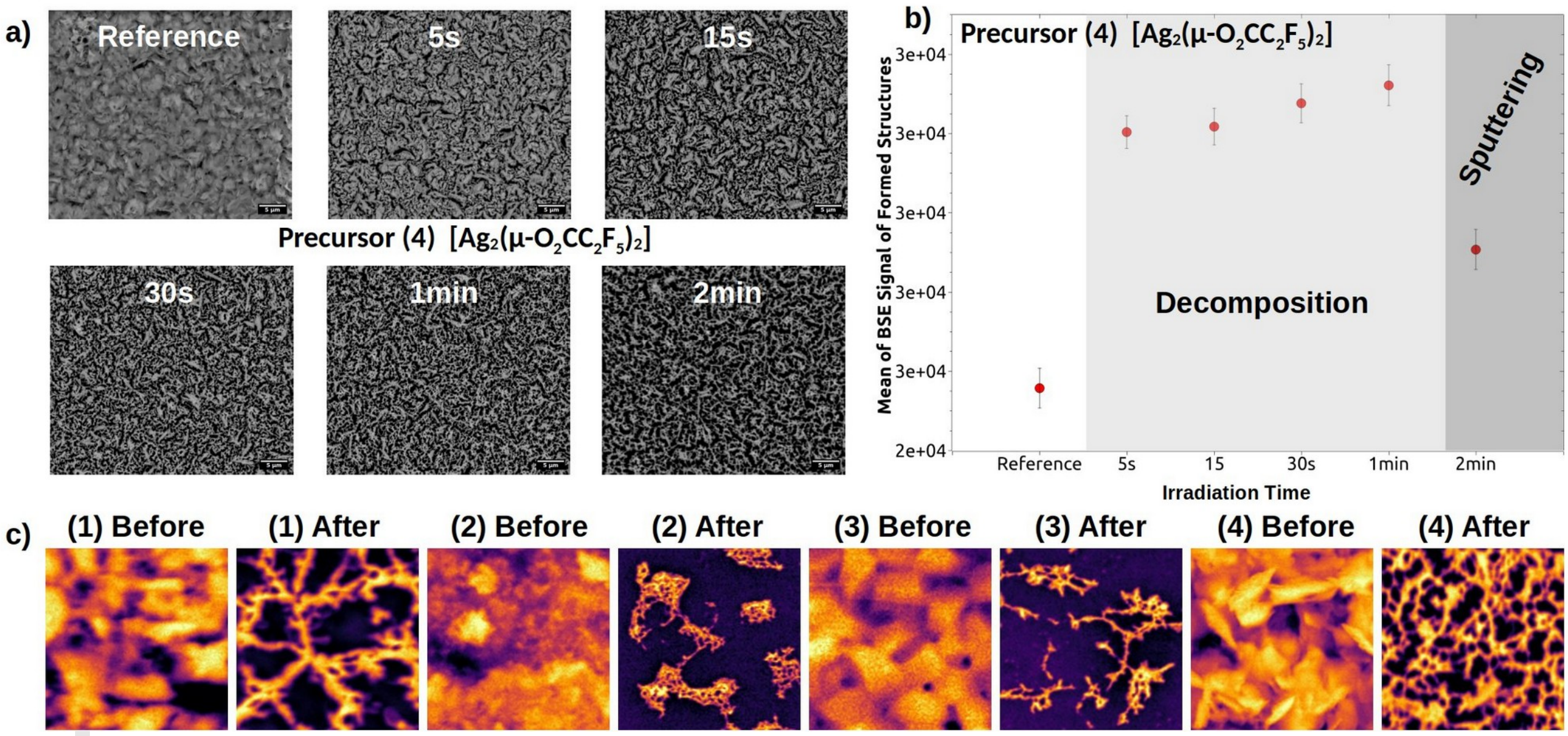

*Figure 2: SEM BSE morphology evolution studies of precursor (4) layer decomposition during gallium FIB irradiation experiments a). SEM BSE intensity (~ atomic number Z) changes of the formed structures, as in a), during gallium FIB irradiation experiments for precursor (4) b). BSE intensity increases during FIB irradiation (metal content increases), the precursor decomposes up to sputtering point at which the formed metal rich structures do not further develop, the sputtering of the structures by FIB gallium ions dominates. SEM BSE morphology before (initial surface) and after gallium FIB decomposition experiments of the layers of precursor (1), (2), (3), (4) respectively c).*

It is seen that for all the precursors the morphology changed significantly, i.e. all the precursors decomposed into metal rich phase. The initially compact films change to a network of interconnected and elongated island-like structures.

In order to determine the chemical composition of resulting morphology, the EDX data were collected in the hyperspectral mode i.e. for each x,y position a full EDX spectrum was collected. The EDX measurements were performed at 20keV electron beam energy.

The EDX data were analyzed firstly by generating netto counts (background subtracted) maps of the elements. Fig. 3a shows the SEM EDX hyperspectral mapping analysis results for the precursor (**4**) after Ga FIB decomposition experiments, including the BSE image and corresponding elemental netto counts maps of C K, O K, F K, Si K, and Ag L lines (see also supporting information Fig.S2-S4 for other precursors). In all cases the maps show that the formed structures are enriched in metal component. In the next step, the EDX data were processed by Machine Learning NMF (non-negative matrix factorization) decomposition as described in details in B.R. Jany et al.[30]. In Fig. 3b-c, the spatial distribution of the individual elements derived from (NMF) decomposition is depicted in the form of loading plots for the 'substrate' and the 'structures'. These plots are shown in color to visually distinguish between the different elements. The substrate layer is shown in blue, while the structures feature is displayed in orange. Additionally, the NMF factors that correspond to the decomposed Electron Dispersive X-ray spectra are shown as well in Fig. 3d. It is evident that the performed NMF decomposition experiments have successfully separated the EDX signal originating from the grown structures from the EDX signal coming from the silicon substrate. Subsequently, these distinct EDX signals were employed to quantify the chemical composition of the structures using an energy-dispersive X-ray EDX ZAF method in a standardless approach.

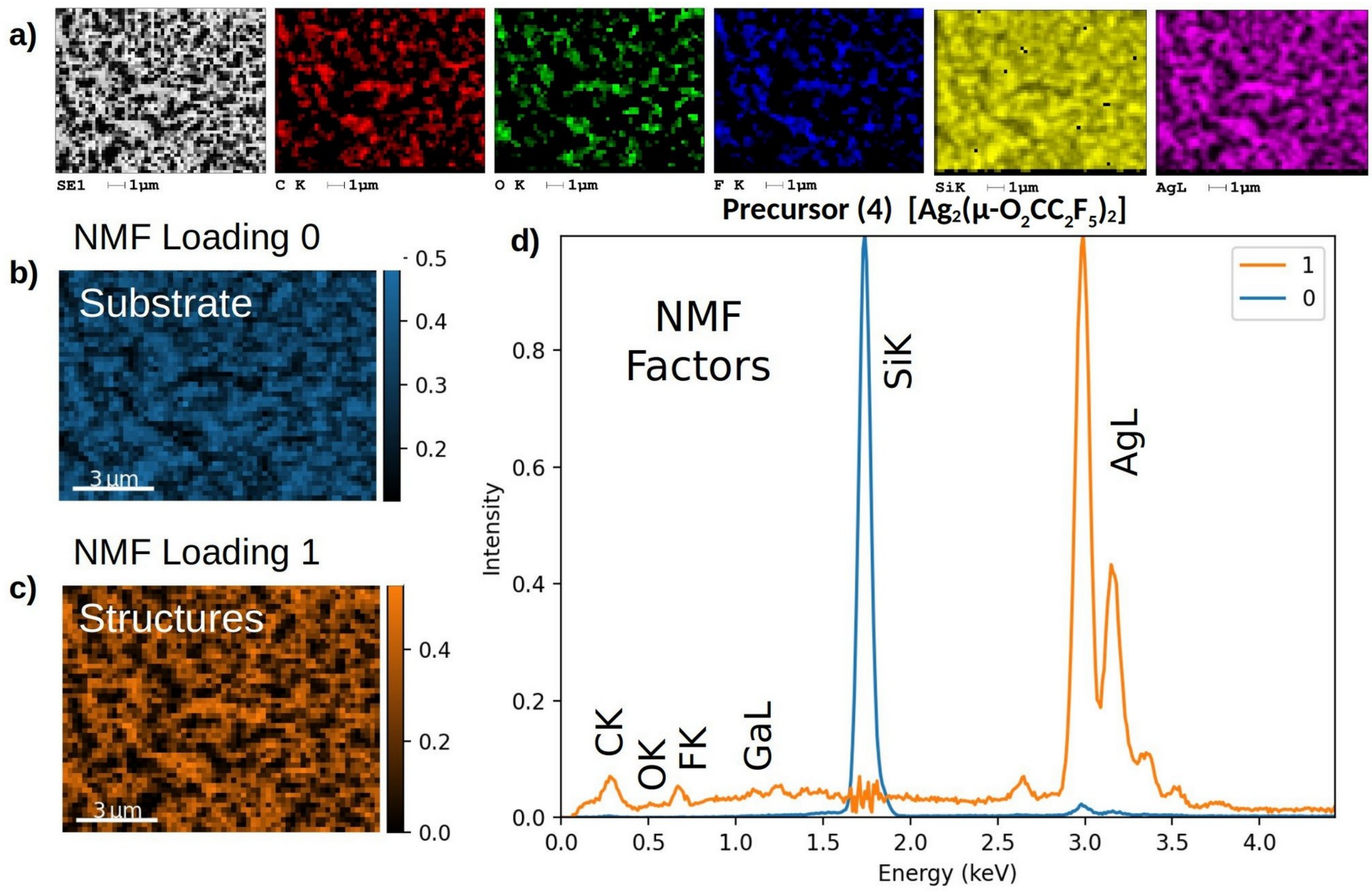

*Figure 3: Results of the SEM EDX hyperspectral mapping analysis of the precursor (4) layer after gallium FIB decomposition experiments, from left BSE image and corresponding elemental netto counts maps of C K, O K, F K, Si K, Ag L lines a). Results of Machine Learning NMF decomposition of the collected SEM EDX hyperspectral data b)-d). NMF loadings showing spatial distribution of the NMF decomposition components b) substrate, c) structures together with NMF factors corresponding to the decomposed EDX signal d). It is seen that the EDX silicon substrate signal is succesfully separated from the signal of the metal rich structures (Si K peak). This allows for the chemical composition quantification via EDX ZAF method.*

## Results and discussion: Chemical composition

The results of the chemical composition of the formed structures are presented in Table 1. The table shows the parameters of gallium ion FIB experiments curried out on deposited precursor layers (**1**−**4**) at their optimal sputtering point, along with scanning electron microscopy and energy dispersive X-ray spectroscopy (EDX) chemical composition analysis in terms of atomic percentages %C, %N, %O, %F, and metal %Cu/%Ag as well as %Ga for the formed structures on the sample surface. For the initial precursor composition please look into Table S2 in Supporting Information. To ensure a fair and accurate comparison between different precursor parameters, it was necessary to take into account the different thicknesses of the precursor layers used in the present study. In order to achieve this, we decided to utilize an ion fluence that had been normalized by the specific height of individual precursor layer Fh=Fluence/(layer height). The fluence [ions/$cm^2$] and fluence per height [ions/$cm^2$/μm] values are also presented in Table 1 as well as the results from previous FEBID experiments for the complexes (**3**) and (**4**)[11, 17]. In the absence of FIBID data for the precursors, the FEBID experiments provide a valuable means of comparison to the gas phase studies. The results allow us to evaluate the performance of the new precursors by correlating ion beam parameters with the chemical composition of the forming structures. Analysis of the data in the table enables us to determine how the potential new precursor is modified under the Ga FIB ion beam exposure. It is seen that for the precursor (**3**) and (**4**), the final metal content obtained under $Ga^+$ FIB irradiation is comparable to that obtained in FEBID experiments.

| **Precursor** | **Parameters of Ga FIB experiments** | | | **SEM EDX content At. % of the formed metal rich structures on the sample surface** | | | | | | **Precursor Score Sp Metal/ (Ga*log(Fh))** |
|---|---|---|---|---|---|---|---|---|---|---|
| | **Fluence [ions/cm$^2$]** | **Fh Fluence/height [ions/cm$^2$/μm]** | **volume/dose [μm$^3$/nC]** | **C** | **N** | **O** | **F** | **Ga** | **Metal** | |
| **(1)** | 1.2*10$^{18}$ | 1.47*10$^{17}$ | 0.0090(0.0054) | 62.57(0.13) | - | 7.73(1.5) | - | 22.22(0.44) | Cu:15.39(0.31) | **0.040** |
| **(2)** | 4.49*10$^{16}$ | 7.45*10$^{15}$ | 0.023(0.014) | 24.5(4.9) | 11.6(2.3) | 6.65(1.3) | 24.2(4.8) | <0.3 | Cu:33.1(0.66) | **6.951** |
| **(3)** | 1.5*10$^{17}$ | 1.06*10$^{17}$ | 0.0030(0.0018) | 22.21(0.89) | - | 10.98(2.2) | 15.18(0.61) | 8.69(0.35) | Cu:42.93(0.86) | **0.290** |
| **(3) FEBID** [11] | - | - | - | 51-5 | - | 2-44 | 44-8 | - | Cu:19-23 | - |
| **(4)** | 1.5*10$^{16}$ | 1.40*10$^{16}$ | 0.21(0.13) | 17.27(1.7) | - | 2.99(1.5) | 15.74(1.6) | 1.44(0.72) | Ag:64.0(1.3) | **2.753** |
| **(4) FEBID** [17] | - | - | - | 20-47 | - | 1-34 | 3-5 | - | Ag:33-76 | - |
| **Pt FIBID** [26,27] | - | - | 0.5 | 24-58 | - | 2-4 | - | 20-28 | Pt:24-46 | - |

*Table 1: Parameters of the final Ga FIB experiments performed (at optimal sputtering point) i.e. Fluence and Fh=Fluence/(layer height) on deposited precursor (**1**), (**2**), (**3**), (**4**) layers together with SEM EDX chemical composition of the formed structures after precursor decomposition. The volume/dose is also estimated for each precursor. The results of the previously performed FEBID experiments are given for the comparison. The longitudinal range of gallium ions (as calculated by SRIM) is equal to 11.6 nm for Cu, 11.0 nm for Ag and 28.6 nm for Si. In the final column, the metric called Precursor Score (Sp) is included, which is calculated as Metal Content divided by [Gallium Content x log(Fh)].*

Furthermore, it can be seen that the gallium content increases with the ion Fluence per height - Fh. Here it is also worth to notice that the gallium ion range in copper and silver is almost the same (the longitudinal range of gallium ions as calculated by SRIM [www.srim.org] is 11.6 nm for copper, 11.0 nm for silver, and 28.6 nm for silicon). Additionally, we calculated the volume-to-dose rate for each studied complex based on the dimensions of the final structures and the applied ion dose. This enables us to compare the ability of studied precursors to form metal-rich structures under a gallium ion FIB irradiation with that of the commonly used Pt precursor (methylcyclopentadienyl) trimethylplatinum(IV) $[Pt(\eta^5\text{-}CpMe)Me_3]$[35,36]. The results indicate that precursors (2) and (4) exhibit performance comparable to the Pt precursor, with precursor (4) being the most comparable. Our ultimate goal was to identify the promising precursor, which would exhibit a high metal content while minimizing gallium accumulation and decomposing efficiently at low ion fluence Fh. Detailed analyses of the chemical composition of the formed structures are presented in Fig. 4. It is evident that the precursor (4) gave the highest metal content among the tested compounds, as observed in the atomic percentage values in Fig. 4a. Fig. 4b presents a visual representation of the ratio of gallium to the total metal content in the final structures. This allows for a comparison of gallium content to other metals within these structures, highlighting variations or trends among the different precursors. The data shows that, in the final metallic structures, precursor (1) results in a significantly higher gallium to other metal ratio compared the other precursors. We also examined the relationship between ion fluence Fh and both the total metal content and the gallium content in the resulting structures. This analysis is illustrated in Fig. 4c and 4d, which show the metal and gallium atomic percentages normalized to ion fluence (Fh). The analysis revealed that precursors (2) and (4) yielded the highest metal content per unit of ion fluence (Fh), indicating that they decomposed most efficiently under Ga ions irradiation. On the other hand, precursor (1) exhibited a higher propensity for gallium absorption during irradiation, as evidenced by its higher gallium to ion fluence (Fh) ratio. By analyzing the ion-beam-induced decomposition of the precursors, which resulted in the formation of structures with varying chemical composition, we were able to assign performance scores to each precursor. To visualize the relationship between three key parameters - gallium content, metal content, and ion Fh - we presented them in a 3-dimensional scatter plot as shown in Fig. 4e. To support the selection of the most suitable precursor, we developed a numeric scoring system a Precursor Score Sp=(Metal Content)/(Galium Content*log(Fh)), as summarized in Table 1. This scoring system prioritizes precursors, which upon decomposition, yield structures with high metal content and low gallium content at the minimal ion fluence Fh, and accurately reflects each precursor candidate's overall performance. This parameter allowed for a quick numerical assessment of precursor performance, helping to identify the most suitable candidates for further investigation or use in FIBID. In this context, the results indicate that precursors (2) and (4) appeared to strike the best balance among all three optimized parameters among the four studied compounds. Finally, to validate our approach, we compared the FEBID metal content data available in the literature (Table 1) with the values obtained using the present layer decomposition methodology (Fig. 4f). The figure shows the average atomic metal content for precursors (3) and (4) measured in FEBID experiments versus the metal content determined using our ion-beam-based decomposition method. It is seen that on the average there is a relation between these two values of the metal content. Independently of the different mechanisms governing layer decomposition under ion beam irradiation and gas-phase deposition in FEBID, our results demonstrate that the proposed method provides valuable insight into the behavior of metal deposition from metalo-organic precursors. We also believe that our methodology could be used as a valuable tool for rapid precursor screening and evaluation.

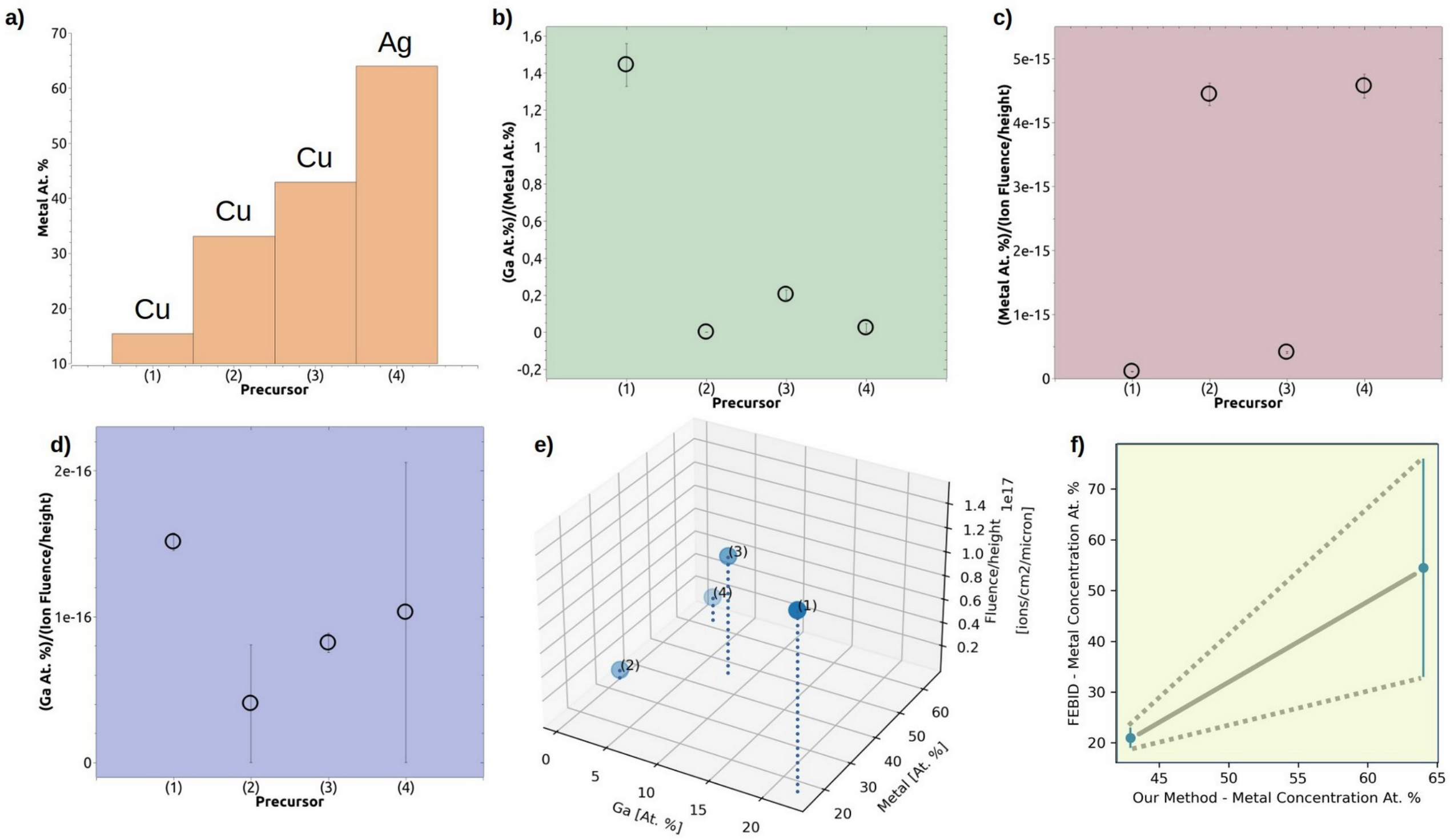

*Figure 4: Investigation of chemical composition (precursor performance) for the studied precursors (1), (2), (3), (4) after gallium FIB experiments. Metal content At. % for the final structures a). It is seen that for the precursor (4) one has the highest metal content. Ratio of gallium to metal for the final structures b). It is seen that precursor (1) accumulates the highest amount gallium in comparison to metal from all precursors. Ratio of metal and gallium At. % to ion fluence Fh c) and d) respectively. It is seen that the precursor (2) and (4) decomposes most easily during FIB experiments producing the highest amount of metal per ion fluence Fh. It is seen that precursor (1) absorbs the most gallium in terms of amount of gallium per fluence during irradiation. A 3-dimensional representation showing the relationship between gallium content, metal content, and ion fluence Fh in a scattered format e). Average metal content in atomic percent obtained for precursors (3) and (4) from the FEBID, versus the metal content of the precursors as determined by our layer decomposition method. A relation between these two values is visible f).*

It's important to note that all of our precursor tests, as well as the final precursor scoring process, were carried out using a straightforward and widely accessible testing method. This approach involved conducting precursor layer tests on SEM microscopy, and utilizing both BSE and EDX analysis techniques. By employing this commonly available methodology, we aimed to ensure that our results would be replicable and relevant to a wide range of potential precursors.

The collected SEM BSE and EDX data together with exemplary python jupyter notebook to analyze EDX hyperspectral data are freely available from Zenodo[37] (https://doi.org/10.5281/zenodo.11354527 ).

## Summary and Conclusions

In this research, we studied the ion-beam-induced decomposition of four Cu or Ag metal-organic precursors (($[Cu_2(\mu\text{-}O_2C^tBu)_4]_n$ (1), $[Cu_2(NH_2(NH=)CC_2F_5)_2(\mu\text{–}O_2CC_2F_5)_4]$ (2), $[Cu_2(\mu\text{-}O_2CC_2F_5)_4]$ (3), and $[Ag_2(\mu\text{–}O_2CC_2F_5)_2]$ (4)) when subjected to gallium focused ion beam irradiation. Individual precursor layers were deposited onto silicon substrates via sublimation and subsequently exposed to gallium focus ion beam (FIB) irradiation. The optimal ion fluence for each precursor was determined by monitoring changes in the backscattered electrons (BSE) signal intensity associated with the evolving surface morphology. The resulting metal-rich surface structures were analyzed using scanning electron microscopy (SEM) and energy dispersive x-ray spectroscopy (EDX) processed by Machine Learning techniques to extract their chemical composition. The study revealed that the silver precursor $[Ag_2(\mu\text{–}O_2CC_2F_5)_2]$ (4) produced the highest overall metal content in the final structures, while the copper precursor (1) resulted in a higher level of gallium incorporation. Copper precursors (2) and (4) demonstrated superior performance in terms of metal yield per unit of ion fluence per height (Fh), indicating that they decomposed more readily compared to the other tested precursors. To evaluate the overall effectiveness of each precursor's, a scoring system called Precursor Score (Sp) was introduced, which incorporates metal content, gallium content, and ion fluence per height required for decomposition. The results showed that precursors (2) and (4) achieved the highest Precursor Scores, indicating their superior ability to balance all three parameters. It is worth to notice that the precursor (4), which is Ag-based, was also tested in FEBID method and gave very good results. To validate our approach, we compared our results with FEBID data, revealing a consistent relationship between the two methods regarding the final metal content. The study highlights the importance of understanding the chemical composition of various potential precursors for producing metal-rich structures using Ga FIB techniques. By employing a straightforward testing methodology, we identified promising carboxylate complexes that could potentially be applied across various fields and applications. Analyzing precursor layer decomposition presents unique challenges compared to gas-phase metal deposition, due to differing underlying processes. However, our method consistently provides valuable insights into the physics governing metal deposition from metal-organic precursors. This approach shows significant potential as a tool for evaluating and selecting promising precursor materials and could accelerate the development of metal-organic precursors specifically tailored for FIB, offering a cost-effective route to novel nanofabrication applications.

**Author Contributions**

A.B.K and K.M synthesized copper and silver complexes, A.B.K prepared the samples for experiments. A.B.K and K.M. selected the precursors for testing and optimized precursor layer deposition process. B.R.J. design the concept of FIB/SEM/BSE experiments and performed all FIB/SEM/BSE/EDX measurements together with data analysis and interpretation. B.R.J. together with I.S.B. prepared the manuscript in consultation with all authors. F.K. and I.B.S. supervised the measurements. All authors have read and agreed to the published version of the manuscript.

## Conflicts of interest

There are no conflicts of interest to declare.

## Acknowledgments

We would like to acknowledge the support of the FIT4NANO Action (COST: CA19140 - Focused Ion Technology for Nanomaterials) for funding the scientific activities and members for sharing their knowledge. This research was supported in part by the Excellence Initiative - Research University Program at the Jagiellonian University in Krakow and by Nicolaus Copernicus University in Torun Grants4NCUStudents 90-SIDUB.6102.66.2022.G4NCUS5.

## Data availability

The collected experimental data are available from Zenodo https://doi.org/10.5281/zenodo.11354527

## References

[1] Li, P.; Chen, S.; Dai, H.; Yang, Z.; Chen, Z.; Wang, Y.; Chen, Y.; Peng, W.; Shan, W.; Duan, H. Recent advances in focused ion beam nanofabrication for nanostructures and devices: Fundamentals and applications. Nanoscale 2021, 13, 1529–1565, doi:10.1039/d0nr07539f.

[2] Barth, S.; Huth, M.; Jungwirth, F. Precursors for direct-write nanofabrication with electrons. J. Mater. Chem. C 2020, 8, 15884–15919

[3] Bruchhaus, L.; Mazarov, P.; Bischoff, L.; Gierak, J.; Wieck, A.D.; Hövel, H. Comparison of technologies for nano device prototyping with a special focus on ion beams: A review; 2017; Vol. 4; ISBN 0000000197762.

[4] De Teresa, J.M.; Orús, P.; Córdoba, R.; Philipp, P. Comparison between focused electron/ion beam-induced deposition at room temperature and under cryogenic conditions. *Micromachines* **2019**, *10*, 1–14, doi:10.3390/mi10120799.

[5] Utke, I.; Swiderek, P.; Höflich, K.; Madajska, K.; Jurczyk, J.; Martinovi´c, P.; Szyma´nska, I.B. Coordination and organometallic precursors of group 10 and 11: Focused electron beam induced deposition of metals and insight gained from chemical vapour deposition, atomic layer deposition, and fundamental surface and gas phase studies. Coord. Chem. Rev. 2022, 458, 213851, doi:10.1016/j.ccr.2021.213851

[6] Fang, C.; Chai, Q.; Lin, X.; Xing, Y.; Zhou, Z. Experiments and simulation of the secondary effect during focused Ga ion beam induced deposition of adjacent nanostructures. Mater. Des. 2021, 209, 10999, 1–11, doi:10.1016/j.matdes.2021.109993.

[7] Córdoba Castillo, R. Ferromagnetic Iron Nanostructures Grown by Focused Electron Beam Induced Deposition. 2014, 71–93, doi:10.1007/978-3-319-02081-5_4.

[8] Matsui, S.; Kaito, T.; Fujita, J. Three-dimensional nanostructure fabrication by focused ion-beam chemical vapor deposition Three-dimensional nanostructure fabrication by focused-ion-beam chemical vapor deposition. 2008, 3181, 3–7, doi:10.1116/1.1319689.

[9] Li, P, et al., Recent advances in focused ion beam nanofabrication for nanostructures and devices: fundamentals and applications, Nanoscale, 2021,13, 1529-1565, doi:10.1039/D0NR07539F

[10] Huang, Y., et al., Microelectromechanical system for in situ quantitative testing of tension–compression asymmetry in nanostructures, Nanoscale Horiz., 2024,9, 254-263, doi:10.1039/D3NH00407D

[11] Sychugov, I., Nakayama, Y., & Mitsuishi, K. (2009). Composition control of electron beam induced nanodeposits by surface pretreatment and beam focusing. Journal of Physical Chemistry C, 113(52), 21516–21519. doi:10.1021/jp9079684

[12] Allen, F.I. A review of defect engineering, ion implantation, and nanofabrication using the helium ion microscope. Beilstein J. Nanotechnol 2021, 12, 633–664, doi:10.3762/bjnano.12.52.

[13] Utke, I.; Michler, J.; Winkler, R.; Plank, H. Mechanical properties of 3d nanostructures obtained by focused electron/ion beam-induced deposition: A review. Micromachines 2020, 11, doi:10.3390/MI11040397.

[14] Indrajith, S., Rousseau, P., Huber, B. A., Nicolafrancesco, C., Domaracka, A., Grygoryeva, K., et al. (2019). Decomposition of iron pentacarbonyl induced by singly and multiply charged ions and implications for focused ion beam-induced deposition. Journal of Physical Chemistry C, 123(16), 10639–10645. doi:10.1021/acs.jpcc.9b00289

[15] Höflich, K.; Hobler, G.; Allen, F.I.; Wirtz, T.; Rius, G.; McElwee-White, L.; Krasheninnikov, A. V.; Schmidt, M.; Utke, I.; Klingner, N.; et al. Roadmap for focused ion beam technologies. Appl. Phys. Rev. 2023, 10, 041311, doi:10.1063/5.0162597.

[16] Berger, L.; Jurczyk, J.; Madajska, K.; Edwards, T.E.J.; Szymańska, I.; Hoffmann, P.; Utke, I. High-Purity Copper Structures from a Perfluorinated Copper Carboxylate Using Focused Electron Beam Induced Deposition and Post-Purification. ACS Appl. Electron. Mater. 2020, 2, 1989–1996, doi:10.1021/acsaelm.0c00282.

[17] Córdoba, R.; Ibarra, A.; Mailly, D.; GuillamÓn, I.; Suderow, H.; De Teresa, J.M. 3D superconducting hollow nanowires with tailored diameters grown by focused He+ beam direct writing. Beilstein J. Nanotechnol. 2020, 11, 1198–1206, doi:10.3762/bjnano.11.104.

[18] Manoccio, M.; Esposito, M.; Passaseo, A.; Cuscunà, M.; Tasco, V. Focused ion beam processing for 3d chiral photonics nanostructures. Micromachines 2021, 12, 1–28, doi:10.3390/mi12010006.

[19] Lee, J.S.; Hill, R.T.; Chilkoti, A.; Murphy, W.L. Surface Patterning. Biomater. Sci. 2020, 553–573.

[20] Court, R.; Jose, S. scan Cr line scan a line is a display is the display of the bottom of the bottom. SPIE 1988, 923, 114–120.

[21] Ivo Utke, Patrik Hoffmann, John Melngailis; Gas-assisted focused electron beam and ion beam processing and fabrication. J. Vac. Sci. Technol. B 1 July 2008; 26 (4): 1197–1276. https://doi.org/10.1116/1.2955728

[22] Felix Jungwirth, Fabrizio Porrati, Daniel Knez, Masiar Sistani, Harald Plank, Michael Huth, and Sven Barth, Focused Ion Beam vs Focused Electron Beam Deposition of Cobalt Silicide Nanostructures Using Single-Source Precursors: Implications for Nanoelectronic Gates, Interconnects, and Spintronics, ACS Applied Nano Materials 2022 5 (10), 14759-14770 DOI: 10.1021/acsanm.2c03074

[23] Martinović, P.; Rohdenburg, M.; Butrymowicz, A.; Sarigül, S.; Huth, P.; Denecke, R.; Szymańska, I.B.; Swiderek, P. Electron-Induced Decomposition of Different Silver(I) Complexes : Implications for the Design of Precursors for Focused Electron Beam Induced Deposition. Nanomaterials 2022, 12, 1687, 1–21.

## References

[24] Berger, L.; Madajska, K.; Szymanska, I.B.; Höflich, K.; Polyakov, M.N.; Jurczyk, J.; Guerra-Nuñez, C.; Utke, I. Gas-assisted silver deposition with a focused electron beam. Beilstein J. Nanotechnol. 2018, 9, 224–232, doi:10.3762/bjnano.9.24.

[25] K. Höflich, J. Jurczyk, Y. Zhang, M. V. Puydinger dos Santos, M. Götz, C. Guerra-Nuñez, J. P. Best, C. Kapusta, I. Utke, Direct Electron Beam Writing of Silver-Based Nanostructures, ACS Appl. Mater. Interfaces 2017, 9, 28, 24071–24077.

[26] J. Jurczyk, K. Höflich, K. Madajska, L. Berger, L. Brockhuis, T. E. J. Edwards, C. Kapusta, I. B. Szymańska, I. Utke, Ligand Size and Carbon-Chain Length Study of Silver Carboxylates in Focused Electron-Beam-Induced Deposition, Nanomaterials 2023, 13(9), 1516; https://doi.org/10.3390/nano13091516.

[27] K. Höflich, J. M. Jurczyk, K. Madajska, M. Götz, L. Berger, C. Guerra-Nuñez, C. Haverkamp, I. B. Szymańska, I. Utke, Towards the third dimension in direct electron beam writing of silver, Beilstein J. Nanotechnol. 2018, 9, 842–849. https://doi.org/10.3762/bjnano.9.78.

[28] X. Guan and R. Yan, Copper-Catalyzed Synthesis of Alkyl-Substituted Pyrrolo[1,2-a]quinoxalines from 2-(1H-Pyrrol-1-yl)anilines and Alkylboronic Acids, Synlett., 2020, 31, 359–362.

[29] A. Butrymowicz-Kubiak, W. Luba, K. Madajska, T. Muzioł and I. B. Szymańska, Pivalate complexes of copper(II) with aliphatic amines as potential precursors for depositing nanomaterials from the gas phase, New J. Chem., 2024, 48, 6232.

[30] Szłyk, E.; Szymańska, I. Studies of new volatile copper(I) complexes with triphenylphosphite and perfluorinated carboxylates. Polyhedron 1999, 18, 2941–2948

[31] E. Szłyk, I. Łakomska, A. Grodzicki, Thermal and spectroscopic studies of the Ag(I) salts with fluorinated carboxylic and sulfonic acid residues, Thermochim. Acta. 223 (1993) 207–212, https://doi.org/10.1016/0040-6031(93)80136-X.

[32] Madajska, K.; Szymańska, I.B. New volatile perfluorinated amidine–carboxylate copper(II) complexes as promising precursors in CVD and FEBID methods. Materials (Basel). 2021, 14, 3145, doi:10.3390/ma14123145.

[33] B.R. Jany, A. Janas, and F. Krok, Retrieving the Quantitative Chemical Information at Nanoscale from Scanning Electron Microscope Energy Dispersive X-ray Measurements by Machine Learning, Nano Letters, Volume 17, Issue 11, 6507-7170 (2017) doi:10.1021/acs.nanolett.7b01789

[34] K. Madajska, Związki koordynacyjne do tworzenia nanostruktur metodą depozycji z fazy gazowej indukowanej zogniskowaną wiązką elektronów, Doctoral dissertation, 2022, (Nicolaus Copernicus University in Torun).

[35] T. Tao, J. S. Ro, J. Melngailis, Z. Xue and H. Kaesz, Focused ion beam induced deposition of platinum, J. Vac. Sci. Technol. B, 8, 1826–9 (1990) doi:10.1116/1.585167

[36] J. Puretz and L. W. Swanson, Focused ion beam deposition of Pt containing films, J. Vac. Sci. Technol. B, 10, 2695–8 (1992) doi:10.1116/1.586028

[37] Jany, B. R. (2024). Data for "Evaluating Metal-Organic Precursors for Focused Ion Beam Induced Deposition through Solid-Layer Decomposition Analysis" [Data set]. Zenodo. https://doi.org/10.5281/zenodo.11354527

# Supporting Information

## Evaluating Metal-Organic Precursors for Focused Ion Beam Induced Deposition through Solid-Layer Decomposition Analysis

B.R. Jany[a*], K. Madajska[b], A. Butrymowicz-Kubiak[b], F. Krok[a], I.B. Szymańska[b]

[a]Marian Smoluchowski Institute of Physics, Faculty of Physics, Astronomy and Applied Computer Science, Jagiellonian University, Lojasiewicza 11, 30348 Krakow, Poland
[b]Faculty of Chemistry, Nicolaus Copernicus University in Toruń, Gagarina 7, 87-100 Toruń, Poland

[*]Corresponding author e-mail: benedykt.jany@uj.edu.pl

### Materials and precursors synthesis

Pivalic acid (99%), $CuCO_3 \cdot Cu(OH)_2$ (>95%), anhydrous acetonitrile (99.8%), and $C_2F_5COOH$ (97%) were purchased from Merck (Saint Louis, MO, USA), absolute ethanol (≥99.8%) – from Honeywell (Charlotte, USA), sodium hydroxide (p.a) – from Avantor (Avantor Performance Materials, Poland S.A.), copper(II) nitrate trihydrate (99%) and $AgNO_3$ (99,9%) – from Chempur (Poland). Pentafluoropropylamidine $C_2F_5C(=NH)NH_2$ (AMDH) (98.7%) was from Apollo Scientific (Stockport, UK). Copper(II) pivalate $[Cu_2(\mu_3\text{-}O_2C^tBu)_2(\mu_2\text{-}O_2C^tBu)_2]_n$ (for simplicity, we use a formula $[Cu_2(\mu\text{-}O_2C^tBu)_4]_n$[1] (**1**)), copper(II) pentafluoropropionate $[Cu_2(\mu\text{-}O_2CC_2F_5)_4]$[2] (**3**), and silver(I) pentafluoropropionate $[Ag_2(\mu\text{–}O_2C_2F_5)_2]$[3] (**4**) were prepared as earlier reported. Based on the synthesis developed by us, the copper(II) amidine−carboxylate complex $[Cu_2(NH_2(NH{=})CC_2F_5)_2(\mu\text{–}O_2CC_2F_5)_4]$[4] (**2**) was synthetized. The Si(111) substrates were purchased from the Institute of Microelectronics and Photonics, Center for Electronic Materials Technology in Warsaw (Lukasiewicz Research Network, Poland).

1 X. Guan and R. Yan, Copper-Catalyzed Synthesis of Alkyl-Substituted Pyrrolo[1,2-*a*]quinoxalines from 2-(1*H*-Pyrrol-1-yl)anilines and Alkylboronic Acids, *Synlett*., 2020, **31**, 359–362.

2 Szłyk, E.; Szymańska, I. Studies of new volatile copper(I) complexes with triphenylphosphite and perfluorinated carboxylates. Polyhedron 1999, 18, 2941–2948.

3 E. Szłyk, I. Łakomska, A. Grodzicki, Thermal and spectroscopic studies of the Ag(I) salts with fluorinated carboxylic and sulfonic acid residues, Thermochim. Acta. 223 (1993) 207–212, https://doi.org/10.1016/0040-6031(93)80136-X.

4 Madajska, K.; Szymańska, I.B. New volatile perfluorinated amidine–carboxylate copper(II) complexes as promising precursors in CVD and FEBID methods. Materials (Basel). 2021, 14, 3145, doi:10.3390/ma14123145.

**Infrared spectra analysis of the original complexes $[Cu_2(\mu\text{-}O_2C^tBu)_4]_n$ (1)[5], $[Cu_2(NH_2(NH=)CC_2F_5)_2(\mu\text{–}O_2CC_2F_5)_4]$ (2)[6], $[Cu_2(\mu\text{-}O_2CC_2F_5)_4]$ (3), and $[Ag_2(\mu\text{–}O_2C_2F_5)_2]$ (4)**

*Table S1: Selected characteristic IR absorption bands ($cm^{-1}$) for $[Cu_2(\mu\text{-}O_2C^tBu)_4]_n$ (1), $[Cu_2(NH_2(NH=)CC_2F_5)_2(\mu\text{–}O_2CC_2F_5)_4]$ (2), $[Cu_2(\mu\text{-}O_2CC_2F_5)_4]$ (3), and $[Ag_2(\mu\text{–}O_2C_2F_5)_2]$ (4).*

| Vibrations | (1) | (2) | (3) | (4) |
|---|---|---|---|---|
| $\nu_{as}COO$ | 1578 and 1530 | 1657 | 1640 | 1603 |
| $\nu_sCOO$ | 1412 | 1418 | 1421 | 1423 |
| $\nu_{as}NH_2$ | − | 3390 | − | − |
| ν(=NH) | − | 3240 | − | − |
| $\delta NH_2$ | − | 1603 | − | − |
| νN=C−N | − | 1510 | − | − |

*Table S2: Initial atomic composition of the precursor layer*

| Precursor | Atomic % | | | | | |
|---|---|---|---|---|---|---|
| | **Cu/Ag** | **C** | **H** | **N** | **F** | **O** |
| **(1) $[Cu_2(\mu\text{-}O_2C^tBu)_4]_n$** | 3,03 | 30,30 | 54,55 | 0,00 | 0,00 | 12,12 |
| **(2) $[Cu_2(NH_2(NH=)CC_2F_5)_2(\mu\text{-}O_2CC_2F_5)_4]$** | 2,94 | 26,47 | 5,88 | 8,82 | 44,12 | 11,76 |
| **(3) $[Cu_2(\mu\text{-}O_2CC_2F_5)_4]$** | 4,76 | 28,57 | 0,00 | 0,00 | 47,62 | 19,05 |
| **(4) $[Ag_2(\mu\text{-}O_2C_2F_5)_2]$** | 9,09 | 27,27 | 0,00 | 0,00 | 45,45 | 18,18 |

[5] A. Butrymowicz-Kubiak, W. Luba, K. Madajska, T. Muzioł and I. B. Szymańska, Pivalate complexes of copper(II) with aliphatic amines as potential precursors for depositing nanomaterials from the gas phase, New J. Chem., 2024, 48, 6232

[6] Madajska, K.; Szymańska, I.B. New volatile perfluorinated amidine–carboxylate copper(II) complexes as promising precursors in CVD and FEBID methods. Materials (Basel). 2021, 14, 3145, doi:10.3390/ma14123145.

**SEM BSE Image Intensity Analysis**

The image analysis, as depicted in Fig. S1, employed a rigorous approach to extract the brightness signal (backscattered electron intensity, BSE) from the metal-rich nanostructures. The initial step involved thresholding using ImageJ/FIJI's default method to effectively separate the metal-rich structures from the underlying silicon substrate. This segmentation enabled the isolation of the regions of interest, thereby facilitating the subsequent analysis.

Subsequently, a histogram of BSE intensities was generated by selectively marking the red areas in the image, which corresponded to the metal-rich nanostructures. This process allowed for the compilation of a comprehensive distribution of BSE intensity values specific to these structures. The mean BSE intensity was subsequently calculated from this data, providing qualitative information about the metal content in the structures since the BSE intensity is proportional to atomic number Z.

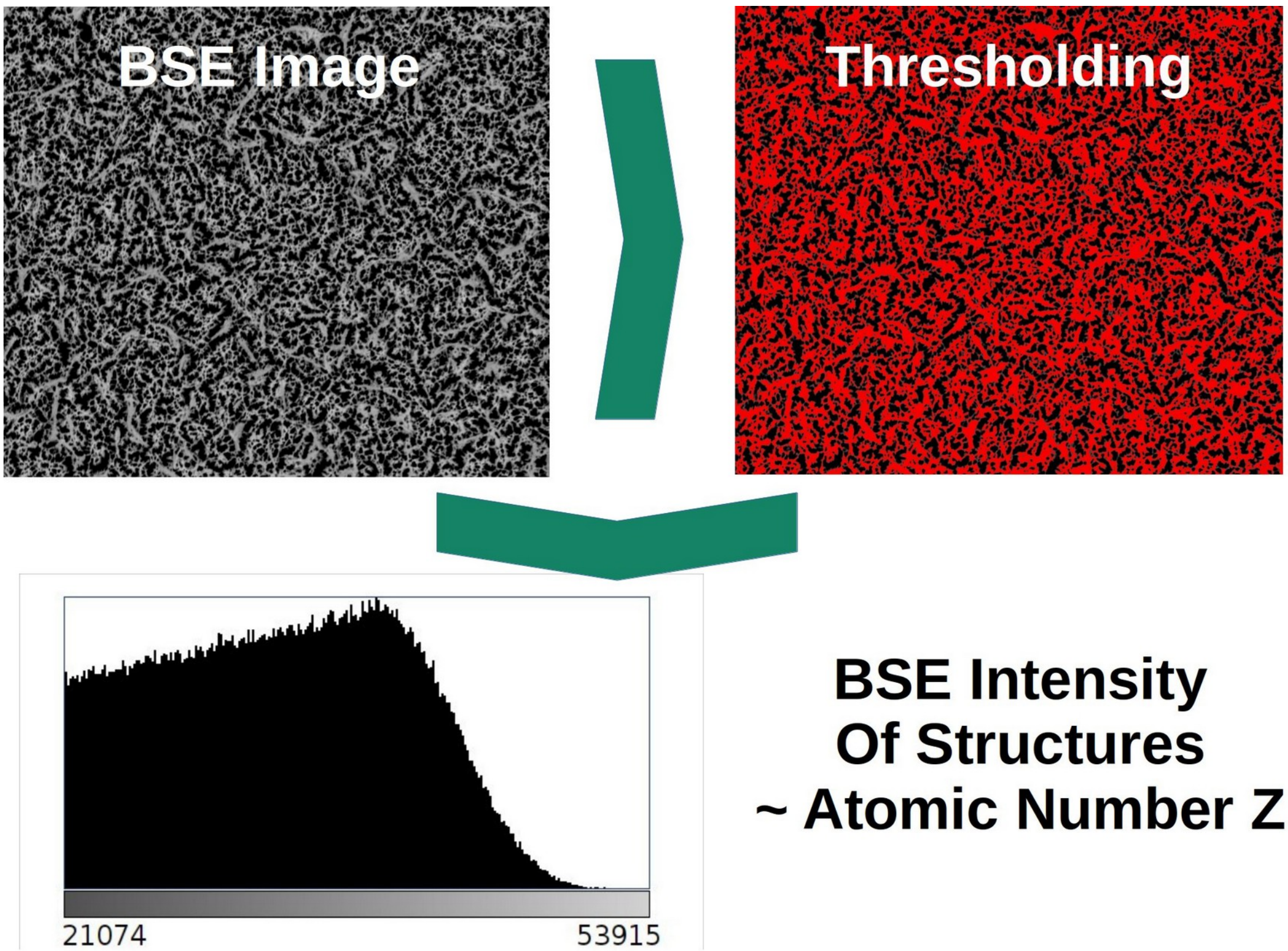

*Figure S1: A comprehensive diagram illustrates the step-by-step workflow for analyzing the brightness signal (backscattered electron intensity, BSE) in images of metal-rich structures obtained after focused ion beam (FIB) irradiation of precursor materials. This schematic representation provides a clear visual overview of the methods employed to extract meaningful information from the BSE images.*

## SEM EDX analysis of the metal rich structures after precursors (1-4) FIB irradiation

**Precursor (1) after FIB Irradiation**

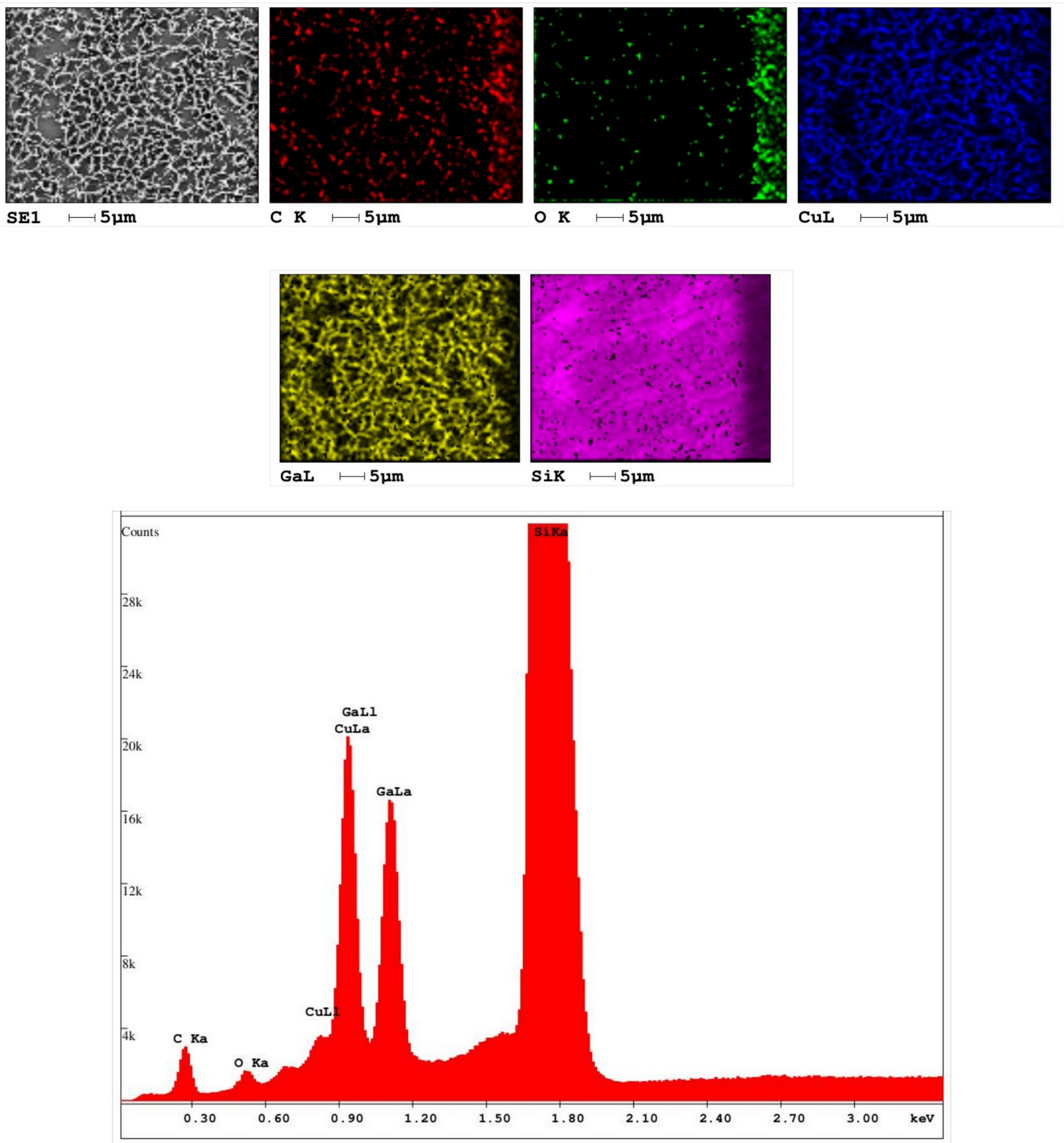

*Figure S2: SEM EDX analysis of the precursor layer (1) $[Cu_2(\mu\text{-}O_2C^tBu)_4]_n$ after FIB irradiation has yielded a set of high-resolution background-subtracted SEM EDX netto-count maps and a comprehensive cumulative EDX spectrum. These maps provide detailed visualizations of the elemental composition and spatial distribution of the sample, while the cumulative spectrum offers quantitative information on the relative abundance of each element present in the material.*

**Precursor (2) after FIB Irradiation**

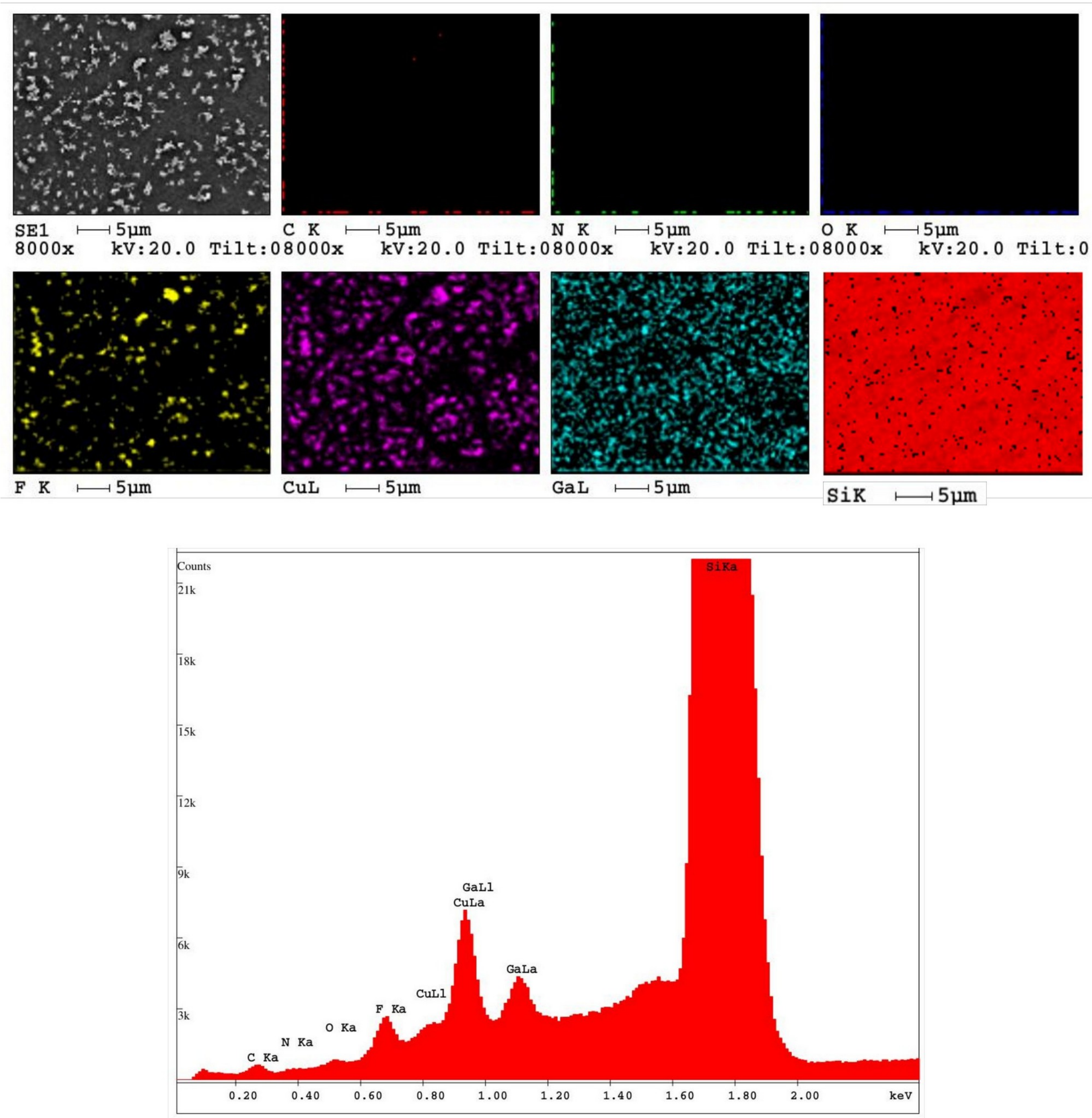

*Figure S3: SEM EDX analysis of the precursor layer (2) [$Cu_2(NH_2(NH{=})CC_2F_5)_2(\mu\text{-}O_2CC_2F_5)_4$] after FIB irradiation has yielded a set of high-resolution background-subtracted SEM EDX netto-count maps and a comprehensive cumulative EDX spectrum. These maps provide detailed visualizations of the elemental composition and spatial distribution of the sample, while the cumulative spectrum offers quantitative information on the relative abundance of each element present in the material.*

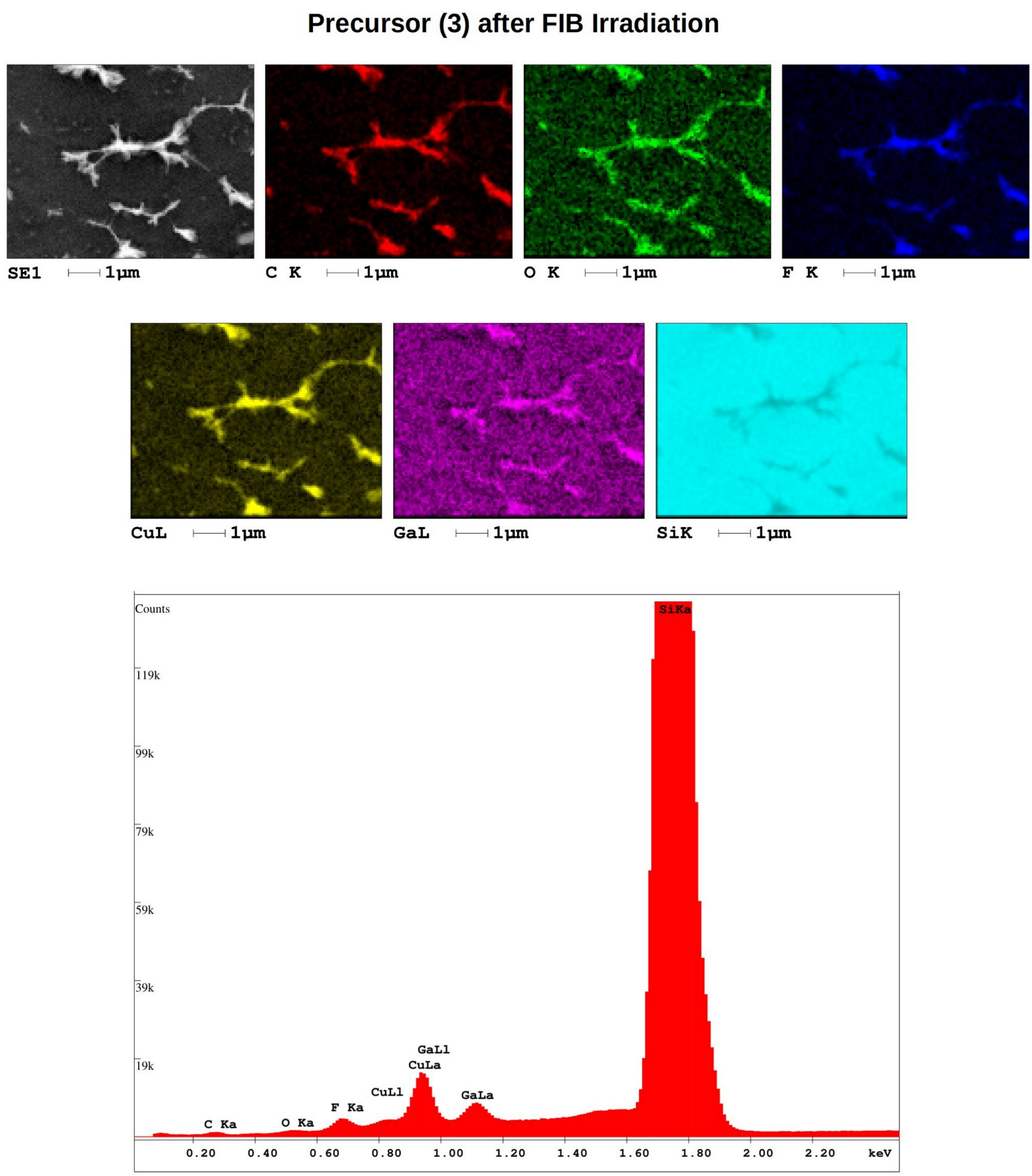

*Figure S4: Results of SEM EDX analysis for the precursor layer (3) $[Cu_2(\mu\text{-}O_2CC_2F_5)_4]$ after FIB irradiation. SEM EDX netto count maps (background subtracted) and corresponding cumulative EDX spectrum. SEM EDX analysis of the precursor material (1) after FIB irradiation has yielded a set of high-resolution background-subtracted SEM EDX net-to-count maps and a comprehensive cumulative EDX spectrum. These maps provide detailed visualizations of the elemental composition and spatial distribution of the sample, while the cumulative spectrum offers quantitative information on the relative abundance of each element present in the material.*

## SEM Electron Beam Irradiation

Following the FIB irradiation of the precursor layer, we also conducted electron beam irradiation experiments under identical conditions and parameters as those employed for ion beam irradiation. Fig. S5 presents SEM BSE images before and after electron beam irradiation, providing a visual comparison of the effects of these two forms of radiation on the precursor layer. This highlights the distinct effects of electron and ion beams on the precursor layer, revealing differences in their transformations.

Notably, despite identical parameters used in this experiment and previous FIB ion beam experiments, the electron beam irradiation does not lead to the decomposition of the precursor layer, differing from the layer transformations observed under FIB ion beam conditions.

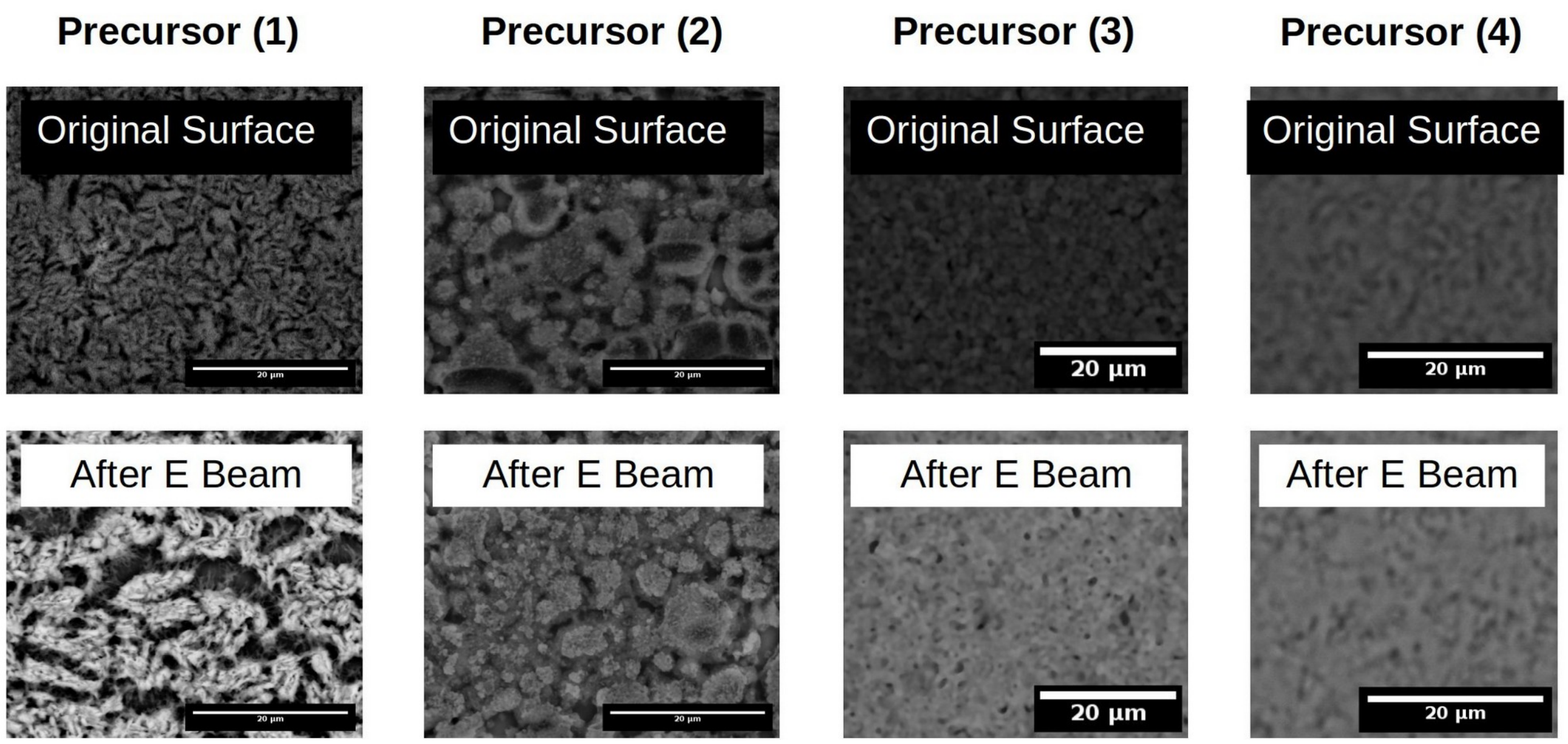

*Figure S5: SEM BSE images are provided for comparison, featuring the original surface morphology of the precursor layers as deposited (upper row) alongside that after electron beam irradiation (lower row). This comparison highlights the distinct effects of these two forms of radiation on the precursor material. Notably, despite identical parameters used in this experiment and previous FIB ion beam experiments, the electron beam irradiation does not lead to decomposition of the precursor layer, differing from the layer transformations observed under FIB ion beam conditions.*